\documentclass[12pt ]{article}
\usepackage{amssymb, amsmath, amsopn, amsthm}
\usepackage{amsmath,bm,amsfonts,amssymb,array,calc,amsthm,rotating}
\usepackage{epsfig}
\usepackage{graphicx}
\usepackage{color}
\usepackage[colorlinks=false]{hyperref}
\def\a{\alpha}
\def\b{\beta}
\def\g{\gamma}
\def\d{\delta}
\def\ep{\epsilon}

\def\m{\mu}
\def\n{\nu}

\def\r{\rho}

\def\s{\sigma}
\def\t{\tau}

\def\G{\Gamma}

\def\O{\Omega}
\def\o{\over}
\def\p{\partial}

\def\p{\partial}

\def\b{\beta}
\def\g{\gamma}

\def\a{\alpha}
\def\G{\Gamma }

\def\e{\varepsilon}

\def\inbar{\,\vrule height1.5ex width.4pt depth0pt}
\def\IC{\relax\hbox{$\inbar\kern-.3em{\rm C}$}}
\def\IN{\relax{\rm I\kern-.18em N}}
\def\IR{\relax{\rm I\kern-.18em R}}
\font\cmss=cmss10 \font\cmsss=cmss10 at 7pt
\def\IZ{\relax\ifmmode\mathchoice
{\hbox{\cmss Z\kern-.4em Z}}{\hbox{\cmss Z\kern-.4em Z}}
{\lower.9pt\hbox{\cmsss Z\kern-.4em Z}} {\lower1.2pt\hbox{\cmsss
Z\kern-.4em Z}}\else{\cmss Z\kern-.4em Z}\fi}

\def\a{\alpha}
\def\b{\beta}
\def\be{\begin{equation}}
\def\beq{\begin{eqnarray}}
\def\g{\gamma}
\def\d{\delta}
\def\ee{\end{equation}}
\def\eeq{\end{eqnarray}}
\def\ep{\epsilon}

\def\m{\mu}
\def\n{\nu}

\def\r{\rho}

\def\s{\sigma}
\def\t{\tau}

\def\G{\Gamma}

\def\O{\Omega}
\def\o{\over}

\def\p{\partial}

\def\p{\partial}

\def\b{\beta}
\def\g{\gamma}

\def\a{\alpha}
\def\G{\Gamma }

\def\e{\varepsilon}

\def\inbar{\,\vrule height1.5ex width.4pt depth0pt}
\def\IC{\relax\hbox{$\inbar\kern-.3em{\rm C}$}}
\def\IN{\relax{\rm I\kern-.18em N}}
\def\IR{\relax{\rm I\kern-.18em R}}

\def\o{\over}

\global\def\draftcontrol{0}

   \def\versionno{Metastable flux configurations}
\catcode`\@=11

\expandafter\ifx\csname
draftcontrol\endcsname\relax\global\def\draftcontrol{0} \fi

{\count255=\time\divide\count255 by 60
\xdef\hourmin{\number\count255} \multiply\count255
by-60\advance\count255 by\time
\xdef\hourmin{\hourmin:\ifnum\count255<10 0\fi\the\count255}}
\def\draftdate{\number\month/\number\day/\number\year\ \ \ \hourmin }

\newcommand\makepapertitle{\par

  \begingroup
    \renewcommand\thefootnote{\@fnsymbol\c@footnote}%
    \def\@makefnmark{\rlap{\@textsuperscript{\normalfont\@thefnmark}}}%
    \long\def\@makefntext##1{\parindent 1em\noindent
            \hb@xt@1.8em{%
                \hss\@textsuperscript{\normalfont\@thefnmark}}##1}%
     \newpage
     \global\@topnum\z@   % Prevents figures from going at top of page.
     \@makepapertitle
     \thispagestyle{empty}\@thanks
  \endgroup
  \setcounter{footnote}{0}%
  \global\let\thanks\relax
  \global\let\makepapertitle\relax
  \global\let\@makepapertitle\relax
  \global\let\@thanks\@empty
  \global\let\@author\@empty
  \global\let\@date\@empty
  \global\let\@title\@empty
  \global\let\title\relax
  \global\let\author\relax
  \global\let\date\relax
  \global\let\and\relax
  \def\version{\let\version\@version\@gobble}
}
\def\@makepapertitle{%
  \newpage
   \ifnum\draftcontrol=1 {}
   \version\versionno
   \vskip 5em%
   \else
   \hfill\hbox to 3cm {\parbox{4cm}{\@pubnum}\hss}%
   \vskip 5em%
   \fi
   \begin{center}%
   \let \footnote \thanks
      {\hskip -0\textwidth \hbox to 1\textwidth%
        {\centerline{\Large\bf{\noindent\@title}}}}%
     \vskip 2em%
     {\normalsize%\large
       \lineskip .5em%
       \begin{tabular}[t]{c}%
         \@author
       \end{tabular}\par}%
     \vskip 1em%
     {\@bstract}%
     \end{center}%
     \vfill
     \@date%
     \vskip 1.5em%
     \noindent
     \rule{12em}{.02em}\par\noindent
     \@email%
   \par
}

\gdef\@pubnum{}
\def\pubnum#1{%
  \gdef\@pubnum{#1}}

\gdef\@bstract{}
\def\Abstract#1{%
  \gdef\@bstract{%
   \parbox{\textwidth-0pc}{%
   \centerline{\bf Abstract}\penalty1000
   \noindent%\abstractfont \baselineskip=12pt
   \renewcommand\baselinestretch{1.0}
   {#1}}}
}

\gdef\@email{}
\def\email#1{%
   \gdef\@email{%
  {\small Email: {\tt #1}}}
}

\def\ps@paper{\let\@mkboth\@gobbletwo%
     \ifnum\draftcontrol=1
        \def\@oddfoot{\hbox to \textwidth{\tiny \versionno \hfil\tiny\draftdate}%
        \hskip -\textwidth \hbox to \textwidth{\hfil\rm\thepage\hfil}}%
     \else\def\@oddfoot{\hbox to \textwidth{\hfil\rm\thepage\hfil}}
     \fi
     \let\@evenfoot\@oddfoot
}
\def\body{\clearpage
          \pagestyle{paper}
        }
\def\@version#1{\ifnum\draftcontrol=1
\typeout{}\typeout{#1}\typeout{}
\vskip3mm\centerline{\hbox{\fbox{\normalsize{\tt DRAFT -- #1 -- }
                   {\draftdate}}}}\vskip3mm
\fi}
\let\version\@version
\long\def\eqlabel#1{\ifnum\draftcontrol=1
                    \tag@false  % there are some problems with multline without this
                    \tag*{(\theequation) \hbox to -0.2cm{\hspace{0cm}\small{#1}\hss}}
                    \refstepcounter{equation}
                    \edef\@currentlabel{\theequation}
                    \ltx@label{#1}          % use old LaTeX \label instead of new definition
                    \else
                    \label{#1}
                    \fi
                    }
\let\st@bibitem\@bibitem
\let\st@lbibitem\@lbibitem
\ifnum\draftcontrol=1
  \def\@bibitem#1{%
    \st@bibitem{#1}\a@@label{#1}\ignorespaces}
  \def\@lbibitem[#1]#2{%
    \st@lbibitem[#1]{#2}\a@@label{#2}\ignorespaces}
  \def\a@@label#1{%
    \gdef\a@lab{\smash{\normalfont\small#1}}
    \ifvmode
      \if@inlabel
        \global\setbox\@labels\hbox{%
          \llap{\a@lab\let\a@lab\relax
                \kern\@totalleftmargin\kern\marginparsep}%
          \box\@labels}%
      \fi
    \fi}
\fi
\usepackage{graphicx}
\usepackage{amsmath,bm,amsfonts,amssymb,array,calc,amsthm,rotating}
\usepackage{epsfig}
\usepackage{graphicx}
\usepackage{color}
\usepackage[colorlinks=false]{hyperref}
\renewcommand\baselinestretch{1.25}
\renewcommand\section{\@startsection {section}{1}{\z@}%
                                   {-3.5ex \@plus -1ex \@minus -.2ex}%
                                   {2.3ex \@plus.2ex}%
                                   {\normalfont\large\bfseries}}
\renewcommand\subsection{\@startsection{subsection}{2}{\z@}%
                                   {-3.25ex\@plus -1ex \@minus -.2ex}%
                                   {1.5ex \@plus .2ex}%
                                   {\normalfont\normalsize\bfseries}}
\renewcommand\subsubsection{\@startsection{subsubsection}{3}{\z@}%
                                   {-3.25ex\@plus -1ex \@minus -.2ex}%
                                   {1.5ex \@plus .2ex}%
                                   {\normalfont\normalsize\it}}
\renewcommand\paragraph{\@startsection{paragraph}{4}{\z@}%
                                   {-3.25ex\@plus -1ex \@minus -.2ex}%
                                   {1.5ex \@plus .2ex}%
                                   {\normalfont\normalsize\bf}}
\renewcommand\subparagraph{\@startsection{subparagraph}{5}{\z@}%
                                   {-1.25ex\@plus -1ex \@minus -.2ex}%
                                   {0ex \@plus .2ex}%
                                   {\normalfont\normalsize\it}}

\numberwithin{equation}{section}

\long\def\@makecaption#1#2{%
  \vskip\abovecaptionskip
  \sbox\@tempboxa{{\bf #1:} #2}%
  \ifdim \wd\@tempboxa >\hsize
    {\small\bf #1:} {\small #2}\par
  \else
    \global \@minipagefalse
    \hb@xt@\hsize{\hfil\box\@tempboxa\hfil}%
  \fi
  \vskip\belowcaptionskip}
\renewcommand*\l@section[2]{%
  \ifnum \c@tocdepth >\z@
    \addpenalty\@secpenalty
    \addvspace{.5em \@plus\p@}%
    \setlength\@tempdima{1.5em}%
    \begingroup
      \parindent \z@ \rightskip \@pnumwidth
      \parfillskip -\@pnumwidth
      \leavevmode \bfseries
      \advance\leftskip\@tempdima
      \hskip -\leftskip
      #1\nobreak\hfil \nobreak\hb@xt@\@pnumwidth{\hss #2}\par
    \endgroup
  \fi}
\renewcommand*\l@subsection{\addvspace{.0em \@plus\p@}\@dottedtocline{2}{1.5em}{2.3em}}
\renewcommand*\l@subsubsection{\addvspace{-.2em \@plus\p@}\@dottedtocline{3}{3.8em}{3.2em}}

\definecolor{refcol}{rgb}{0.2,0.2,0.8}
\definecolor{eqcol}{rgb}{.6,0,0}
\definecolor{purple}{cmyk}{0,1,0,0}

\gdef\@citecolor{refcol} \gdef\@linkcolor{eqcol}
\def\colorlinkspurple{\gdef\@urlcolor{purple}}
\def\colorlinksblue{\gdef\@urlcolor{blue}}
\def\colorlinksred{\gdef\@urlcolor{red}}

\def\revise#1       {\raisebox{-0em}{\rule{3pt}{1em}}%
                     \marginpar{\raisebox{.5em}{\vrule width3pt\
                     \vrule width0pt height 0pt depth0.5em
                     \hbox to 0cm{\hspace{0cm}{%
                     \parbox[t]{4em}{\raggedright\footnotesize{#1}}}\hss}}}}

\def\ee           {{\it e}}

\def\sqr#1#2{{\vcenter{\vbox{\hrule height.#2pt
 \hbox{\vrule width.#2pt height#1pt \kern#1pt
 \vrule width.#2pt}\hrule height.#2pt}}}}

  \def\a{\alpha}
  \def\b{\beta}

\catcode`\@=12

\begin{document}

\title{Metastable Flux Configurations and de Sitter Spaces}

%\pubnum{%
%arXiv:yymm.nnnn}
\date{June 2007}

\author{\\[.5cm]Katrin Becker, Yu-Chieh Chung
and Guangyu Guo \
\\[.2cm]\it Department of Physics,
Texas A\&M University, \\ \it College Station, TX 77843, USA\\
[1.5cm]}

\Abstract{We derive stability conditions for the critical points of
the no-scale scalar potential governing the dynamics of the complex
structure moduli and the axio-dilaton in compactifications of type
IIB string theory on Calabi-Yau three-folds. We discuss a concrete
example of a $T^6$ orientifold. We then consider the
four-dimensional theory obtained from compactifications of type IIB
string theory on non-geometric backgrounds which are mirror to rigid
Calabi-Yau manifolds and show that the complex structure moduli
fields can be stabilized in terms of $H_{RR}$ only, {\it i.e.} with
no need of orientifold projection. The stabilization of all the
fields at weak coupling, including the axio-dilaton, may require to
break supersymmetry in the presence of $H_{NS}$ flux or corrections
to the scalar potential.}

\email{kbecker,ycchung,guangyu@physics.tamu.edu}

%\enlargethispage{1.5cm}

\makepapertitle

\body

\version\versionno

\vskip 1em

%\tableofcontents

%\newpage

\newpage

\section{Introduction}

There are many ways to obtain models of particle phenomenonlogy
using string theory. A good starting point is to construct a model
with ${\cal N}=1$ supersymmetry in four dimensions. One can obtain
such models, for example, by compactifying M-theory on
G$_2$-holonomy manifolds, F-theory on Calabi-Yau four-folds or type
II theories on Calabi-Yau orientifolds.  It is a beautiful fact that
these models have a moduli space of vacuum states. However, concrete
predictions can only be made if the mechanism which picks the vacuum
state of string theory can be identified. By including fluxes as
background fields the continuous ambiguity associated with the
vacuum expectation values of the moduli fields is replaced by a
discrete freedom associated with the choice of flux numbers.
However, the number of possible vacuum states is still enormous and
it has been argued to built a whole landscape of solutions. However,
most of these string theory backgrounds have flat directions and the
number of solutions with all moduli stabilized is very limited.

Stabilizing all the scalar fields associated with a Calabi-Yau
compactification of string theory at weak coupling is a particularly
hard problem. In the context of compactifications of type IIB string
theory on a Calabi-Yau orientifold, for example, one of the fields
which is conventionally stabilized using fluxes is the axio-dilaton.
This removes the arbitrariness associated with the vacuum
expectation value of this field. However at the same time this means
that the string coupling constant is no longer a parameter whose
value we can freely choose but it is determined in terms of fluxes.
Weak coupling can then only be achieved if some flux numbers can be
chosen to be large. But taking such a limit and making the moduli
fields heavy is difficult, and it has been conjectured in ref.
\cite{bbw} that it is actually impossible. This situation may change
once supersymmetry is broken and as a result it is very important to
determine the properties of flux configurations leading to stable
critical points of the scalar potential while breaking
supersymmetry. This is the aim of the present paper.

Lets illustrate this idea in the interesting example of the KKLT
model \cite{kklt}. Here complex structure moduli and the
axio-dilaton acquire an expectation value due to perturbative fluxes
while preserving an ${\cal N}=1$ supersymmetry. Non-perturbative
corrections to the superpotential cause the radial modulus $\rho$ to
become heavy compared to the AdS cosmological constant while the
masses of the complex structure moduli will generically be of the
order to the inverse AdS length and may not be heavy enough to be
considered stabilized \cite{bbw}. This situation changes once these
vacua are lifted to dS spaces. According to ref. \cite{kklt} this
can be achieved by assuming the presence of an anti-D3 brane which
contributes a factor
\begin{equation}
\Delta V\sim {1\over ({\rm Im} \rho)^3},
\end{equation}
to the scalar potential. Once this contribution is taken into
account the potential for the radial modulus displays a metastable
minimum at which the scalar potential takes a positive value.
Moreover, the masses of the moduli are of the order to the AdS scale
which can be much larger than the dS scale and as a result after
supersymmetry is broken all the moduli fields can turn out to be
heavy enough.

Adding anti-D3 branes is one way to uplift the potential to positive
values. Following the argument of ref. \cite{ss}, it should also be
possible to obtain a potential contribution resembling the one
resulting from anti-D3 branes by considering flux configurations for
which ${\cal D}_I W \neq 0$ for some $I$. From the no-scale form of
the potential it follows that such a contribution is positive and
it's dependence on $\rho$ is precisely equal to the one originating
from anti-D3 branes. This makes it a natural alternative to the KKLT
model. Since ${\cal D}_IW\neq 0$ the flux can no longer be imaginary
self-dual (ISD) but will acquire an imaginary anti-self dual (IASD)
component. Requiring that the scalar potential is critical in the
complex structure and axio-dilaton directions imposes conditions on
the fluxes which we will derive in the present paper while the
radial modulus is not stabilized.

We then consider the four-dimensional theory obtained from
compactifications of type IIB string theory on backgrounds which are
mirror to rigid Calabi-Yau manifolds, {\it i.e.} non-geometric
backgrounds with no K\"ahler structure. In this case case the flux
induced superpotential does depend explicitly on all scalar fields,
{\it i.e.} the complex structure moduli and the axio-dilaton. Mirror
symmetry implies that on the type IIB side the K\"ahler potential
for the axio-dilaton differs from the conventional one obtained from
dimensional reduction \cite{bbw}. This fact enables us to find a
scalar potential which stabilizes all the complex structure moduli
in terms of RR fluxes only while requiring no orientifold charge.
However the axio-dilaton is not fixed and slides off to weak
coupling. The axio-dilaton could be stabilized if $H_{NS}$ is taken
into account and supersymmetry is broken to render the scalar fields
heavy enough. Another possibility is to take perturbative
corrections to the K\"ahler potential and non-perturbative
corrections to the superpotential into account \cite{bbw}.

The organization of this paper is as follows. In section 2 we
consider geometric compactifications of type IIB string theory on
Calabi-Yau three-folds. We derive the conditions imposed on the flux
configurations to lead to stable critical points of the scalar
potential in the complex structure and axio-dilaton directions. We
explicitly show that the critical points do correspond to minima of
the potential by computing the Hessian matrix. We illustrate the
idea in the example of a torus orientifold. In section 3 we consider
the four-dimensional theory obtained from compactifications of type
IIB strings on mirrors of rigid Calabi-Yau manifolds. We find a
scalar potential which stabilizes all the complex structure moduli
in terms of RR fluxes only while requiring no orientifold charge. We
discuss several possibilities to stabilize the axio-dilaton at weak
coupling. We then end with some conclusions and speculations.

\section{Type IIB string theory compactified on Calabi-Yau three-folds}

In this section we discuss geometric compactifications of type IIB
strings and analyze the critical points of the scalar potential. To
set up the notation we start in subsection 2.1 deriving the form of
the scalar potential following closely ref. \cite{gkp}. In
subsection 2.2 we derive the conditions to obtain a critical point
of the potential. In subsection 2.3 we explicitly check that the
critical points correspond to minima by computing the Hessian
matrix. In subsection 2.4 we present a concrete example.

\subsection{The scalar potential}

Our starting point is the low-energy effective action of type IIB
strings in the ten-dimensional Einstein frame
\begin{eqnarray}
\label{action} S_{IIB} & = &  {1\o 2{\kappa}_{10}^2}\int
d^{10}x\sqrt {-g}\left[R-{\p_M\t\p^M\bar \t\o 2(\mbox{Im}
\t)^2}-{G\cdot\overline
G\o 12 \mbox{Im}\t}-{\tilde F^2_{(5)}\o 4\cdot 5!}\right] \label{action}\\
 &  &
-{1\o 8i \kappa_{10}^2}\int{C_{(4)}\wedge G\wedge \overline G\o
\mbox{Im}\t}+S_{loc}.  \nonumber
\end{eqnarray}
Here the axio-dilaton $\tau$ is written in terms of the RR scalar
$C_{(0)}$ and the dilaton $\phi$ according to
\begin{equation}
\t=C_{(0)}+ie^{-\phi},
\end{equation}
and
\begin{equation}
\tilde F_{(5)}=F_{(5)}-{1\o 2}C_{(2)}\wedge H_{NS}+{1\o 2}
B_{(2)}\wedge H_{RR},
\end{equation}
where $H_{RR}$ and $H_{NS}$ are the two three-forms with potentials
$C_{(2)}$ and $B_{(2)}$ respectively and $G=H_{RR}-\tau H_{NS}$. The
condition that $\tilde F_{(5)}$ is self-dual should be imposed by
hand. The Bianchi identity for the five-form field is
\begin{equation}
\label{bianchi} d\tilde F_{(5)}=H_{NS}\wedge
H_{RR}+2\kappa_{10}^2T_3\r_3^{loc}.
\end{equation}
By integrating the Bianchi identity over the internal manifold
${\cal M}_6$, we get
\begin{equation}
\label{tadpole} {1\o(2\pi)^4\a'^{2}}\int_{{\cal M}_6} H_{NS}\wedge
H_{RR}+Q_3^{loc}=0,
\end{equation}
where we have used the relation $2\kappa_{10}^2T_3=(2\pi)^4\a'^{2}$.
This identity means the sum of the D3 charges from background fields
and localized sources vanishes. From Eq. (\ref{action}) one obtain
the four-dimensional scalar potential by dimensional reduction
\begin{equation}
\label{Scalar potential} V={1\o
24\kappa^2_{10}(\mbox{Im}\r)^3}\int_{{\cal M}_6}d^6y \sqrt g {G\cdot
\overline G\o \mbox{Im}\t}-{i\o
4\kappa^2_{10}(\mbox{Im}\r)^3}\int_{{\cal M}_6}{G\wedge\overline
G\o\mbox{Im}\t}
\end{equation}

By using the flux induced superpotential \cite{superpotential} which
is explicitly given by
\begin{equation}
\label{gvw} W=\int_{{\cal M}_6} G \wedge \Omega,
\end{equation}
and the K\"ahler potential
\begin{equation}
\label{Kahler} {\cal K}= -3\log [-i(\r-\bar \r)]- \log [-i(\t-\bar
\t)]-\log [-i\int_{{\cal M}_6}\O\wedge\overline\O],
\end{equation}
where $\rho$ is the radial modulus, the scalar potential
(\ref{Scalar potential}) can be transformed into the standard ${\cal
N}=1$ supergravity form
\begin{equation}
\label{Supergravity1} V=e^{\cal K}\left(g^{a\bar b}{\cal D}_aW {\cal
D}_{\bar b}\overline W-3|W|^2\right)
\end{equation}
where $a$ and $b$ label all moduli and the axio-dilaton.
 Because the superpotential is independent
of $\rho$ the scalar potential takes the no-scale form
\begin{equation}
\label{Supergravity2} V=e^{\cal K}F_I \bar F^I ,
\end{equation}
where $I$ and $J$ label the complex structure moduli and the
axio-dilaton. Here and in the following we will be using the
notation of \cite{denef+douglas}
\begin{equation}
\label{notation} F_I = {\cal D}_I W, \qquad Z_{IJ} = {\cal D}_I
{\cal D}_J W, \qquad U_{IJK} = {\cal D}_I {\cal D}_J {\cal D}_K W,
\end{equation}
and indices are raised using the inverse of the K\"ahler metric
$g_{I\bar J}=\p_I\p_{\bar J}{\cal K}$.

\subsection{Critical points of the scalar potential}

The scalar potential will be critical in the complex structure and
axio-dilaton directions if the first derivatives vanish, {\it i.e.}
if
\begin{equation}\label{ai}
\partial_I V = e^{\cal K} \left( Z_{IJ} \bar F^J + F_I \overline W\right) =0.
\end{equation}
One, but not the most general, solution of this condition is given
by flux configurations satisfying $F_I=0$. Using the explicit
expression for the superpotential we have
\begin{equation}
F_i = \int_{{\cal M}_6} G\wedge \chi_i\qquad {\rm and } \qquad
F_\tau=-{1\o \t- \bar \t}\int_{{\cal M}_6}\overline G\wedge \O,
\end{equation}
where $\chi_i$ is the basis of harmonic $(2,1)$ forms and with lower
case indices $i,j$ we label the complex structure moduli only. This
implies that the non-vanishing components of $G$ can lie in the
$(0,3)$ or $(2,1)$ directions. In other words, $G$ is ISD, $\star
G=i G$. Moreover, this critical point is stable because the scalar
potential (\ref{Supergravity2}) is positive semi-definite and at the
critical points the potential vanishes.

In the following we would like to find the most general solution of
Eq.  (\ref{ai}). We start by rewriting Eq. (\ref{ai}) in the form
\begin{equation}\label{aii}
\begin{split}
& Z_{\tau \tau} \bar F ^\tau +Z_{\tau j} \bar F^j + F_\tau \overline
W=0,\\
& Z_{i \tau} \bar F^\tau + Z_{ij} \bar F^j + F_i \overline W=0.\\
\end{split}
\end{equation}
Note that
\begin{equation}
Z_{ij}=\kappa_{ijk} { \int_{{\cal M}_6} G\wedge \overline\chi^k\over
\int _{{\cal M}_6} \Omega \wedge \overline \Omega}, \quad Z_{\tau i}
=-{1\over \tau -\bar \tau} \int_{{\cal M}_6} \bar G \wedge \chi_i,
\quad Z_{\tau \tau} = 0.
\end{equation}
A simple computation (we include the details in an appendix) shows
that the first condition in Eq. (\ref{aii}) is equivalent to
\begin{equation}\label{aiii}
\int_{{\cal M}_6} G \wedge \star G=0,
\end{equation}
while the second condition leads to
\begin{equation}
\label{condition}
    ( B \bar B_k+A \bar A_k)\int_{{\cal M}_6} \O\wedge\overline
  \O+\kappa_{ijk}A^i B^j=0.
\end{equation}
Here we introduced the Hodge decomposition
\begin{equation}
G=A\O+A^i\chi_i+\bar B^{\bar i}\bar \chi_{\bar i}+\bar B\overline \O
\end{equation}
and $\kappa_{ijk}$ are the Yukawa couplings. From here we conclude
that the potential for the scalars (\ref{Supergravity2}) does not
have a critical point for an arbitrary choice of flux. Only if Eq.
(\ref{aiii}) and Eq. (\ref{condition}) are satisfied can we find a
critical point in all directions except the size. This is not always
possible. If $H_{NS}=0$, for example, then the dilaton cannot be
stabilized since the only non-vanishing contribution to the dilaton
potential comes from the overall factor
 $e^{\cal K}$. As a result no critical point exists since Eq.
 (\ref{aiii}) is violated.

It is not difficult to see that all flux combinations can lead to
critical points of the potential except if $G$ is given by a
combination of the following components
\begin{equation}
G_{(3,0)}+G_{(0,3)}, \qquad G_{(3,0)}+G_{(2,1)}, \qquad
G_{(3,0)}+G_{(0,3)}+G_{(2,1)},
\end{equation}
or their complex conjugates. A flux of the form
$G_{(3,0)}+G_{(0,3)}$, for example, is easily seen to violate the
condition (\ref{aiii}).

Among the possible flux combinations leading to critical points of
the scalar potential only a flux lying in the $(2,1)$ or $(1,2)$
directions preserves supersymmetry. The $(2,1)$ component obviously
preserves supersymmetry, as it satisfies
\begin{equation}
{\cal D}_IW={\cal D}_\r W=0.
\end{equation}
However a flux in the $(1,2)$ direction also preserves supersymmetry
if accompanied by a change in the sign of the tadpole due to fluxes.
The reason for this is that type IIB supergravity in ten dimensions
is invariant under the change of sign of all RR fluxes. Changing the
signs of RR fields replaces $G$ by $-\bar G$ and as a result a flux
lying in the (2,1) direction should lead to the same physics as a
flux in the (1,2) direction. The (1,2) component does satisfy the
conventional supergravity constraint ${\cal D}_I\widetilde W={\cal
D}_\r \widetilde W=0$, but with a superpotential given by
\begin{equation}
{\widetilde W}=\int_{M_6} \overline G\wedge \O.
\end{equation}
The derivation of this superpotential will be discussed in appendix
B. The two superpotentials $W$ and $\widetilde W$ are related to
each other by a CPT transformation. Any other flux components
satisfying Eq. (\ref{aiii}) and (\ref{condition}) will not preserve
supersymmetry and lead to a positive cosmological constant or
vanishing cosmological constant if only a $(3,0)$ (or $(0,3)$)
component is turned on. On the other hand, due to the no-scale
structure of the potential the radial modulus cannot be stabilized.

\subsection{The Hessian matrix}

The no-scale potential is positive definite. As a result solutions
which lead to a vanishing potential at the critical point $V_\star$
are necessarily stable. However, we are interested in solutions for
which $V_\star
>0$ and as a result we have to check the stability of the solutions.
In order to determine if the critical points are stable we compute
the Hessian matrix $H$. It turns out that it only has positive
eigenvalues which means that the critical points are minima in the
complex structure and axio-dilaton directions. Indeed, the second
derivatives of the scalar potential are given by
\begin{eqnarray}
\label{seconderivative}
\p_I\p_J V&=&e^{\cal K}\left(U_{IJK}\bar F^K +2Z_{IJ}\bar W \right)\\
\p_I\p_{\bar J}V&=&e^{\cal K}(g_{I\bar J}F_K\bar F^K-R_{I\bar
JK}^{\;\;\;\;\;\;\;L}F_L\bar F^K+2F_I\bar F_{\bar J} +Z_{IL}\bar
Z_{\bar J\bar K}g^{L\bar K} +g_{I\bar J}|W|^2)\nonumber
\end{eqnarray}
The critical points will be stable if
\begin{equation}
d \Sigma^2=H_{\alpha \beta}dw ^\a dw^\b\ge 0,
\end{equation}
where $w^\alpha$ labels all coordinates, {\it i.e.} $\alpha$ and
$\beta$ label the axio-dilaton, complex structure moduli and their
complex conjugates. Using formulas which are explicitly presented in
appendix A we obtain
\begin{equation}
\label{Hessian} d \Sigma^2=e^{\cal K}g^{\g \s} \left( Z_{\a \g}
\overline Z_{\b \s}dw^\a dw^\b+g^{\t\bar \t}U_{\a \g \t }\overline
U_{\b
 \s \bar \t}dw^\a dw^\b\right)
\end{equation}
where $U_{\a\g\s}= {\cal D}_\a {\cal D}_\g{\cal D}_\s W$ and $Z_{\a
\g}= {\cal D}_\a{\cal D}_\g W$  are the generalization of $U_{IJK}$
and $Z_{IJ}$. As a result the hessian matrix is positive
semi-definite and the critical points correspond to minima.

\subsection{An example}

In this section we describe a concrete example in terms of a type
IIB orientifold compactification. This example is closely related to
examples discussed in refs. \cite{ss} and \cite{kst}. We will be
following their notation. Let $x^i$ and $y^i$, for $i=1,2,3$ be the
six real coordinates on $T^6$. These coordinates are subjected to
the periodic identifications $x^i\equiv x^i+1$ and $y^i\equiv
y^i+1$. The complex structure is parameterized by complex parameters
$\t^{ij}$, and
\begin{equation}
z^i=x^i+\t^{ij}y^j,
\end{equation}
are global holomorphic coordinates. The explicit orientifold is
$T^6/\O R(-1)^{F_L}$, where $R$ is the involution which changes the
sign of all torus coordinates, $R:(x^i,y^i)\to -(x^i,y^i)$. The
holomorphic three-form is
\begin{equation}
\O=dz^1\wedge dz^2\wedge dz^3,
\end{equation}
and the metric is
\begin{equation}
ds^2=dz^id\bar z^{\bar i}.
\end{equation}
We choose the following orientation
\begin{equation}
\int_{T^6} dx^1\wedge dx^2\wedge dx^3\wedge dy^1\wedge dy^2\wedge
dy^3=1,
\end{equation}
and the basis of $H^3(T^6,\mathbb{Z})$:
\begin{eqnarray}
\a_0 &=&dx^1\wedge dx^2\wedge dx^3 \nonumber \\\a_{ij}
&=&{1\o2}\e_{ilm}dx^l\wedge dx^m\wedge dy^j ,  \;\;\;\;1\leq i,j\leq
3 \nonumber\\ \b^{ij} &=&-{1\o2}\e_{jlm}dy^l\wedge dy^m\wedge dx^i,
\;\;\;1\leq i,j\leq 3\nonumber\\ \b^0 &=& dy^1\wedge dy^2\wedge
dy^3\label {basis}
\end{eqnarray}
which satisfies $ \int_{T^6}\a_I\wedge \b^J=\d_I^J$. The fluxes can
be expanded in this basis
\begin{eqnarray}
{1\o(2\pi)^2\a'}H_{RR}&=&a^0\a_0+a^{ij}\a_{ij}+b_{ij}\b^{ij}+b_0\b^0\label{flux}\\
{1\o(2\pi)^2\a'}H_{NS}&=&c^0\a_0+c^{ij}\a_{ij}+d_{ij}\b^{ij}+d_0\b^0.
\nonumber
\end{eqnarray}
Here we take $a^0,a^{ij},b_0,b_{ij},c^0,c^{ij},d_0,d_{ij}$ to be
even integers, so that all the O3-planes are of the standard type
and the issues regarding flux quantization discussed in ref.
\cite{fp} can be avoided. In this case, the total number of
O3-planes is 64 and each plane has D3-brane charge $-1/4$. For
simplicity we only turn on the diagonal components of the flux, so
that we can set the off-diagonal components of $\t^{ij}$ equal to
zero at the critical points. This condition can be imposed by
restricting to an enhanced symmetry locus on the moduli space of the
$T^6$ \cite{ss}. For example, we will consider configurations which
are symmetric under
\begin{eqnarray}
R_1: (x^1,x^2,x^3,y^1,y^2,y^3) &\to&
(-x^1,-x^2,x^3,-y^1,-y^2,y^3)\nonumber\\
R_2:(x^1,x^2,x^3,y^1,y^2,y^3)&\to& (x^1,-x^2,-x^3,y^1,-y^2,-y^3)
\end{eqnarray}
Only the diagonal components of the complex structure $\t^{ij}$, and
the three forms $\a_0,\a_{ii},\b^0,\b^{ii}$ are preserved under
these symmetries, so that the only non-vanishing flux components are
$a^0$, $a^{ii}$, $b_0$, $b_{ii}$ and $c^0$, $c^{ii}$, $d_0$,
$d_{ii}$. We are left with 3 non-vanishing complex moduli and the
axio-dilaton.

To use the conditions (\ref{aiii}) and (\ref{condition}) which we
derived in subsection 2.2, we need to transform the scalar potential
(\ref{Scalar potential}) into the standard ${\cal N}=1$ supergravity
formula (\ref{Supergravity1}). For tori having a general complex
structure the result is complicated (see for example \cite{aft} and
\cite{ss}). However for tori with diagonal complex structure, we can
express the scalar potential in the form
\begin{equation}
\label{scalar potential2} V=e^{\cal K}\left(\sum_{i,j=1}^3 g^{i\bar
j}{\cal D}_{\t_i}W\overline {{\cal D}_{\t_j}W} +g^{\t\bar \t }{\cal
D}_\t W\overline{{\cal D}_{\t}W}\right),
\end{equation}
with superpotential (\ref{gvw}) and  ``K\"ahler potential",
\begin{equation}
\label{kahler2}{\cal K}= -3\log [-i(\r-\bar \r)]- \log [-i(\t-\bar
\t)]-\log [i(\t_1-\bar \t_1)(\t_2-\bar \t_2)(\t_3-\bar \t_3)],
\end{equation}
where we used $\t_i$ to replace $\t^{ii}$. Before we proceed we have
one more comment. Generally we can only set $\t^{ij}=0$ (for $i\neq
j$), after computing the first derivative of the scalar potential
(\ref{Scalar potential}), but on the symmetric locus, the
criticality conditions $\p_{\t^{ij}}V=0$ (for $i\neq j $) are
automatically satisfied. As a result we can set $\t^{ij}=0$ (for
$i\neq j$) at the beginning of the computation and only deal with
the conditions $\p_{\t^{ii}}V=0$. However, when computing the second
derivatives we can not set $\t^{ij}=0$ before we differentiate, as
there are non-vanishing terms of the form $\p^2_{\t^{ij}}V$, which
will disappear if we set $\t^{ij}=0$ (for $i\neq j$) at the
beginning.

Next we consider a flux in the (2,1)+(1,2) direction, so the
conditions (\ref{condition}) and (\ref{aiii}) take the form
\begin{equation}
\label{condition1} \kappa_{ijk}A^jB^k=0\qquad {\rm and}\qquad
g_{i\bar j}A^i\bar B^{\bar j}=0.
\end{equation}
Since we are working with a torus we set $\kappa_{123}=1$ and one
solution to the above condition is
\begin{equation}
\label{condition2} A^3=B^3=0,\qquad A^1 B^2=-B^1 A^2,\qquad {A^1\bar
B^1\o (\mbox{Im}\t_1)^2}+{A^2\bar B^2\o (\mbox{Im}\t_2)^2}=0.
\end{equation}
For the concrete torus orientifold we are considering the tadpole
cancelation condition takes the form
\begin{equation}
\label{tadpole1} {i\o 2\mbox{Im}\t
(2\pi)^4\a'^2}\int_{T^6}G\wedge\overline G=32.
\end{equation}

In the following we will present a concrete solution of Eq. (\ref
{condition2}). For simplicity we redefine the parameters according
to
\begin{equation}
A^i=-2i\mbox{Im}\t_i\mbox{Im}\t \tilde A ^i,\qquad {\rm and} \qquad
\bar B^{\bar i}=2i\mbox{Im}\t_i\mbox{Im}\t \tilde {\bar B}^{\bar i}
\end{equation}
and drop the tilde in the following. The conditions (\ref
{condition2}) and (\ref{tadpole1}) can be written as
\begin{equation}
\label{condition3} A^1 B^2=-B^1 A^2, \qquad B^1\bar B ^1=B^2\bar
B^2,\qquad (A^2\bar A^2-B^2\bar
B^2)\mbox{Im}\t\prod_{i=1}^3\mbox{Im}\t_i=4
\end{equation}
and the non-vanishing components of $H_{RR}$ and $H_{NS}$ are
\begin{equation}\label{fluxcomponents}
\begin{split}
& a^0= -\mbox{Im}[\bar \t(A^1+A^2+\bar B^1+\bar B^2)]\\
& a^{11}= -\mbox{Im}[\bar \t(A^1\bar\t_1+A^2\t_1+\bar B^1\t_1+\bar B^2\bar \t_1)]\\
& a^{22}=-\mbox{Im}[\bar \t(A^1\bar\t_2+A^2\t_2+\bar B^1\t_2+\bar B^2\bar \t_2)]\\
& a^{33}=-\mbox{Im}[\bar \t(A^1\bar\t_3+A^2\t_3+\bar B^1\t_3+\bar B^2\bar \t_3)]\\
& b_{0} =-\mbox{Im}[\bar \t(A^1 \bar \t_1\t_2\t_3+A^2
\t_1\bar\t_2\t_3+
\bar B^1 \t_1\bar\t_2\bar\t_3+\bar B^2\bar\t_1\t_2\bar\t_3)]\\
& b_{11}=\mbox{Im}[\bar \t(A^1\t_2\t_3+A^2\bar\t_2\t_3+\bar
B^1\bar\t_2\bar\t_3+\bar B^2\t_2\bar\t_3)]\\
& b_{22}=\mbox{Im}[\bar \t(A^1\bar \t_1\t_3+A^2\t_1\t_3+\bar B^1
\t_1\bar\t_3+\bar B^2\bar\t_1\bar\t_3)]\\
& b_{33}=\mbox{Im} [\bar \t(A^1\bar \t_1\t_2+A^2 \t_1\bar\t_2+\bar
B^1 \t_1\bar\t_2+\bar
B^2\bar\t_1\t_2)]\\
& c^0=-\mbox{Im}[A^1+A^2+\bar B^1+\bar B^2] \\
   &     c^{11}= -\mbox{Im}[A^1\bar\t_1+A^2\t_1+\bar B^1\t_1+\bar B^2\bar \t_1]\\
& c^{22}=-\mbox{Im}[A^1\bar\t_2+A ^2\t_2+\bar B^1\t_2+\bar B^2\bar \t_2]\\
& c^{33}=-\mbox{Im}[A^1\bar\t_3+A^2\t_3+\bar B^1\t_3+\bar B^2\bar \t_3]\\
& d_{0} =-\mbox{Im}[A^1 \bar \t_1\t_2\t_3+A^2 \t_1\bar\t_2\t_3+
\bar B^1 \t_1\bar\t_2\bar\t_3+\bar B^2\bar\t_1\t_2\bar\t_3]\\
& d_{11}=\mbox{Im}[A^1\t_2\t_3+
A^2\bar\t_2\t_3+\bar B^1\bar\t_2\bar\t_3+\bar B^2\t_2\bar\t_3]\\
& d_{22}=\mbox{Im}[A^1\bar \t_1\t_3+A^2\t_1\t_3+\bar B^1
\t_1\bar\t_3+\bar B^2\bar\t_1\bar\t_3]\\
& d_{33}=\mbox{Im}[A^1\bar \t_1\t_2+A^2 \t_1\bar\t_2+\bar B^1
\t_1\bar\t_2+\bar B^2\bar\t_1\t_2].\\
\end{split}
\end{equation}

Usually one starts with certain flux numbers and then determines the
values of moduli fields. Here we solve the inverse problem, namely,
we start with the value of the moduli and determine the flux numbers
which stabilize the moduli at the given values. To solve Eq.
(\ref{condition3}) using even flux numbers (\ref{fluxcomponents}) we
use the ansatz
\begin{equation}
\mbox{Im}\t=4,\qquad \mbox{Im}\t_1=\mbox{Im}\t_2=\mbox{Im}\t_3=1
\end{equation}
So one solution of Eq. (\ref{condition3}) is
\begin{equation}
A^1=-3i,\qquad A^2=3 i,\qquad \bar B^1=2+2i,\qquad \bar B^2=2+2i
\end{equation}
From Eq. (\ref{fluxcomponents}), we can explicitly compute the flux
numbers and obtain
\begin{equation}
\begin{split}
& (a^0,a^{11},a^{22},a^{33}) =(16, -24, 24,
16),
\\
& (b_0,b_{11},b_{22},b_{33}) =(16, 0, 0, -16)\\
& (c^0,c^{11},c^{22},c^{33}) =(-4, 0, 0, 4),\\
& (d_0,d_{11},d_{22},d_{33}) =(4, 6, -6, 4)\\
\end{split}
\end{equation}
which are all even integrals.

\section{Type IIB mirrors of type IIA strings compactified on rigid
Calabi-Yau three-folds}

In this section we would like to generalize the previous analysis to
type IIB theories which arise as mirrors of type IIA models
compactified on rigid Calabi-Yau three-folds, {\it i.e.} with
$h_{2,1}=0$. On the type IIB side these correspond to models with
$h_{1,1}=0$ and consequently are not ordinary Calabi-Yau manifolds
since a K\"ahler form is missing but can nevertheless be described
using conformal field theory techniques. Here we will be interested
in the properties of the resulting four-dimensional theories which
contain $h_{2,1}+1$ four-dimensional ${\cal N}=1$ chiral superfields
originating from the complex structure moduli and the axio-dilaton.
The number of these fields will in general be reduced if we consider
an orientifold projection.

It has been shown in ref. \cite{bbw} that for compactifications of
type IIB strings on backgrounds with no K\"ahler structure the
K\"ahler potential for the axio-dilaton and the complex structure is
\begin{equation}\label{bi}
{\cal K} = - 4 \log \left[-i (\tau - \bar \tau) \right]-
\log\left[-i \int \Omega \wedge \bar \Omega \right],
\end{equation}
which differs by a subtle factor 4 from the conventional K\"ahler
potential for the axio-dilaton. This unconventional factor 4 has the
consequence that supersymmetric flux configurations are no longer
required to be ISD \cite{bbw}. The K\"ahler potential (\ref{bi})
also causes the scalar potential to display new and interesting
properties. In order to illustrate this imagine one considers a real
three-form flux, {\it i.e.} a flux configuration with $H_{NS}=0$.
Then
\begin{equation}
W=W_{RR}=\int H_{RR}\wedge \Omega,
\end{equation}
and the scalar potential can be written in the form
\begin{equation}\label{aaiii}
V = e^{\cal K} \left(g^{i \bar j } D_i W_{RR} \overline{  D_{ j}
W_{RR}}+\mid W_{RR} \mid^2 \right),
\end{equation}
which is positive definite and depends non-trivially on the complex
structure. If
\begin{equation}\label{aai}
\partial_i V=0 \qquad {\rm for  } \qquad i=1,\dots,h_{2,1},
\end{equation}
the potential is critical in all the complex structure directions.
So for example, one solution of Eq. (\ref{aai}) is given by
\begin{equation}\label{aaii}
H_{RR}=a \left( \Omega + \bar \Omega \right),
\end{equation}
where $a$ is some real constant. This equation determines the
complex structure moduli. Indeed, it turns out that this is nothing
else than the equation defining a rank 1 attractor which is well
known from black hole physics. Eq. (\ref{aaii}) can, for example, be
explicitly solved in the large complex structure limit as has been
shown by Shmakova in ref. \cite{Shmakova:1996nz} (see also ref.
\cite{Moore:1998pn}). These critical points are stable since the
only non-vanishing entries of the Hessian matrix are
\begin{equation}
\partial_{\bar i} \partial_j V= 2 e^{\cal K} g_{\bar i j} |
W_{RR}|^2.
\end{equation}
The scalar potential (\ref{aaiii}) has been studied before in the
literature in the context of non-supersymmetric attractors (for a
partial list of references on non-supersymmetric attractors see
\cite{attrac}). In particular, the critical points of the potential
are the solutions of
\begin{equation}
H_{RR}= 2 {\rm Im} \left[ e^{{\cal K}_{cs}}\left( \Omega\bar W -
\bar F^i \chi_i \right)\right] ,
\end{equation}
subjected to the constraint
\begin{equation}
Z_{ij}\bar F^j+2F_i\bar W=0
\end{equation}
which can be written as
\begin{equation}
2 F_i \bar W \int \O\wedge \overline\O+ \kappa_{ijk} \bar F^j \bar
F^k=0.
\end{equation}
Moreover, these critical points are stable since the Hessian matrix
written in terms of\footnote{Here the indices $\alpha,\beta$ label
the complex structure moduli and their complex conjugates.}
\begin{equation}
 d \Sigma^2=2e^{\cal K} \left( g^{\g \s}Z_{\a \g}
\overline Z_{\b \s}dw^\a dw^\b+\bar F_\a F_\b dw^\a dw^\b\right),
\end{equation}
is positive definite. In this form the critical points correspond
non-supersymmetric attractor points as described in ref.
\cite{Kallosh:2005ax}. This indicates that within a non-geometric
model with $h_{1,1}=0$ the proposal of ref. \cite{ss} leads to an
interesting new class of backgrounds in which all the complex
structure moduli can be stablized in terms of RR fluxes only with no
need of negative energy sources like orientifold planes.

Using the solution (\ref{aaii}) shows that the potential at the
minimum satisfies
\begin{equation}
V_\star >0,
\end{equation}
if $a\neq 0$ so the external space is dS. However, before we can
conclude that supersymmetry is spontaneously broken by the solution
(\ref{aaii}) we should take into account the dependence on the
axio-dilaton arising from the overall factor $e^{\cal K}\sim ({\rm
Im} \tau)^{-4}$. This factor causes the potential to slope to zero
at infinity so a supersymmetric state is gained back at infinity and
as it stands the theory has no ground state at all. Here (as in
\cite{bbw}) we will simply assume that perturbative corrections to
the K\"ahler potential and non-perturbative corrections to the
superpotential could achieve this stabilization and lead to a
metastable ground state.

In order to stabilize the axio-dilaton using perturbative fluxes the
only possibility is to use a non-vanishing $H_{NS}$ flux. By
including RR and NS three-form fluxes one obtains a four-dimensional
superpotential which does depend non-trivially on all moduli fields.
Any geometric compactification would lead to a superpotential which
is independent of the K\"ahler moduli and consequently the radial
modulus would slide off to infinity. As a result even in the absence
of any type of corrections moduli stabilization may be possible
within the non-geometric model by including all possible fluxes.
Moreover, in order to obtain moduli fields which are heavy enough we
may have to break supersymmetry \cite{bbw}. But note that once the
NS flux is non-vanishing the scalar potential is no longer positive
definite and it is not obvious that supersymmetry breaking vacua,
and in particular the phenomenologically interesting vacua leading
to a positive cosmological constant, exist. As an illustrative toy
example lets consider a non-geometric model with $h_{2,1}=0$, {\it
i.e.} a model with only one massless scalar field, the axio-dilaton,
with a K\"ahler potential
\begin{equation}\label{aax}
{\cal K} = - 4 \log \left[ - i (\tau - \bar \tau) \right],
\end{equation}
and a superpotential
\begin{equation}
W=W_{RR}-\tau W_{NS},
\end{equation}
where $W_{RR}$ and $W_{NS}$ are constants. The condition for
unbroken supersymmetry has one solution only
\begin{equation}
\tau = {1\over W_{NS} \bar W_{NS}} \left[{\rm Re}(\bar W_{NS}
W_{RR}) + 2 i {\rm Im}(\bar W_{NS}  W_{RR} )\right].
\end{equation}
However, it is not difficult to see that the scalar potential is
also critical if
\begin{equation}
\tau = {1\over W_{NS} \bar W_{NS}} \left[{\rm Re}(\bar W_{NS}
W_{RR}) -{ i\over 2} {\rm Im}(\bar W_{NS}  W_{RR} )\right],
\end{equation}
which leads to $D_\tau W\neq 0$ so that supersymmetry is broken.
Moreover, the scalar potential at the minimum is negative so that
the external space is AdS. As a result supersymmetry breaking
critical points of the potential do exist even though in this case
they lead to an AdS space. However, it is interesting that a single
four-dimensional chiral field with a K\"ahler potential of the form
(\ref{aax}) avoids the no-go theorem of ref. \cite{Lebedev:2006qc}
(see also \cite{nogo}) according to which dS or Minkowski space
vacua with a broken supersymmetry are never possible in a theory
with a single chiral field for any superpotential if the K\"ahler
potential is ${\cal K}=-n \log \left[ -i(\tau - \bar \tau)\right]$
with $1\leq n \leq 3$. As a result stable dS vacua are no longer
excluded. It will be very interesting to see if by considering a
`realistic' model with a non-vanishing number of complex structure
moduli fields stable critical points of the potential at which
supersymmetry is broken can be found. We leave this topic for future
research.

\section{Conclusion and speculations}

In this paper we have analyzed stability conditions of the no-scale
scalar potential determining the dynamics of the complex structure
moduli and the axio-dilaton in geometric flux compactification of
type IIB strings to four dimensions. In order to obtain critical
points which do not preserve supersymmetry an essential ingredient
is the appearance of IASD flux components. But not any IASD flux is
allowed. Fluxes have to satisfy the conditions (\ref{aiii}) and
(\ref{condition}) to stabilize all fields except the K\"ahler
structure moduli.

Searching for the critical points of the scalar potential obtained
from compactifications of type IIB strings on mirrors of rigid
Calabi-Yau three-folds we discovered a fascinating and unexpected
analogy to black hole physics and, in particular, to
non-supersymmetric attractors. This mapping was possible because of
the peculiarities of the axio-dilaton K\"ahler potential in the
non-geometric setting derived recently in ref. \cite{bbw}. The
similarities between the attractor mechanism and flux
compactifications have been known for some time (see for example
ref. \cite{Moore:2004fg}) but an explicit mapping of the scalar
potentials is new. Complex structure moduli stabilization can be
achieved in terms of $H_{RR}$ only, {\it i.e.} in terms of a real
three-form. It will be interesting to further explore if lessons
learned from black hole physics can be used to discover properties
of flux vacua with a small and positive cosmological constant.

Moreover, compactifications of type IIB strings on mirrors of rigid
Calabi-Yau manifolds lead to a flux induced superpotential which
depends non-trivially on all scalar fields even in the absence of
any non-perturbative effects once the RR and NS three-form fluxes
are included. This leads to the interesting possibility that
perturbative fluxes alone may stabilize all moduli fields once
supersymmetry is broken.

\vskip 1cm

\noindent {\bf Acknowledgment}

\noindent We would like to thank Melanie Becker, Jason Kumar, Ergin
Sezgin, Eva Silverstein, Cumrun Vafa and Johannes Walcher for
valuable discussions and communications. This work was supported in
part by NSF grants PHY-0505757 and the University of Texas A\&M.
K.B. would like to thank the Galileo Galilei Institute for
Theoretical Physics and the CERN theory division for hospitality and
partial financial support during the completion of this work.

\section{Appendix A}

In this appendix, we present the details of some of the computations
presented in this paper. To set up our notation we start by
reviewing a few basic formulas regarding Calabi-Yau manifolds
\cite{candelas}. On a Calabi-Yau three-fold, there exists a unique
harmonic (3,0) form $\O$, whose first derivatives satisfy
\begin{equation}
{\p\O\o \p z^i}=K_i\O+\chi_i \qquad  {\rm and} \qquad {\p\O\o \p
\bar z^i}=0
\end{equation}
where $\chi_i$ is an harmonic (2,1) form. The K\"ahler potential on
the complex structure moduli space is
\begin{equation}
\label{kahler3} K_{cs}=-\log[-i\int\O\wedge\overline\O].
\end{equation}
As is easy to check
\begin{equation}
\p_i K_{cs}=-K_i\qquad {\rm and} \qquad g_{i\bar j}=\p_i\p_{\bar
j}K_{cs}=-{\int \chi_i\wedge \bar \chi_{\bar j}\o \int \O\wedge
\overline\O}.
\end{equation}
One important property of the (3,0) form $\O$ is that it is
undefined up to multiplication by a holomorphic function $f(z)$
\begin{equation}\label{aaxxx}
\O\to f(z) \O.
\end{equation}
Under (\ref{aaxxx}) the K\"ahler potential transforms as
\begin{equation}
K_{cs}\to K_{cs}-\log f(z) - \log \bar f(\bar z),
\end{equation}
which leaves the metric on moduli space invariant. For convenience,
we can define a gauge covariant derivative
\begin{equation}
\chi_i={\cal D}_i \O=\p_i\O+\p_i K_{cs}\O,
\end{equation}
and thus under the K\"ahler transformation, it transforms according
to ${\cal D}_i \O\to f{\cal D}_i \O$, {\it i.e.} $ \chi_i\to
f\chi_i$. One can also generalize the definition of the covariant
derivative to other quantities which transform like
\begin{equation}
\Psi^{(a,b)}\to f^a\bar f^b \Psi^{(a,b)}
\end{equation}
under the K\"ahler transformation. In this case the covariant
derivatives take the form
\begin{eqnarray}
{\cal D}_i \Psi^{(a,b)}& =&(\p_i+a\p_i K_{cs})\Psi^{(a,b)}\nonumber\\
{\cal D}_{\bar j} \Psi^{(a,b)} &=&(\p_{\bar j}+b\p_{\bar j}
K_{cs})\Psi^{(a,b)}.
\end{eqnarray}
The partial derivatives $\p_i$ and $\p_{\bar i}$ are to be replaced
by ordinary covariant derivatives $\nabla_i$, $\nabla_{\bar j}$ when
acting on tensors. It is easy to see that under K\"ahler
transformations
\begin{equation}
{\cal D}_{i} \Psi^{(a,b)}\to f^a\bar f^b {\cal D}_{i}
\Psi^{(a,b)}\qquad {\rm and}\qquad {\cal D}_{\bar j} \Psi^{(a,b)}\to
f^a\bar f^b {\cal D}_{\bar j} \Psi^{(a,b)}.
\end{equation}
We also require
\begin{equation}[{\cal D}_i, {\cal D}_{\bar
j}]\O=-g_{i\bar j}\O,\qquad {\rm and}\qquad  {\cal D}_k g_{i\bar
j}=0.
\end{equation}
Using the above formulas, we can get the results quoted in the table
below
\begin{equation}
\label{table}
\begin{tabular}{|c|c|}\hline
  \hline
 \emph{Derivatives of the basis} & \emph{Spans}\\ \hline
 $ \O$ & (3,0) \\
  ${\cal D}_i \O=\chi_i$ & (2,1)\\
  ${\cal D}_i\chi_j={1\o \int \O\wedge \overline\O}\kappa_{ij}^{\;\;\;\;\bar k}\bar \chi_{\bar k}$ & (1,2)
  \\
  ${\cal D}_i\bar\chi_{\bar j}=g_{i\bar j}\overline \O$  & (0,3) \\
  ${\cal D}_i \overline \O=0$ &\\\hline
\end{tabular}
\end{equation}
where the Yukawa couplings are defined as
\begin{equation}
\kappa_{ijk}=\int \O\wedge {\cal D}_i{\cal D}_j{\cal D}_k\O.
\end{equation}

The above results are the tools needed to compute the first
derivative of scalar potential (\ref{Supergravity2}). Because the
scalar potential is invariant under K\"ahler transformation, {\it
i.e.} $a=b=0$, we can transform the ordinary derivatives into
covariant derivatives
\begin{equation}
\p_I V={\cal D}_I V=e^{\cal K} \left( Z_{IJ} \bar F^J + F_I
\overline W\right)
\end{equation}
with the notation (\ref{notation}). To obtain an explicit expression
for $\p_I V=0$, we need to compute a few quantities,
\begin{eqnarray}
F_i &=& {\cal D}_i W=\int_{{\cal M}_6} G\wedge \chi_i\nonumber\\
F_\tau &=& {\cal D}_\t W=\p_\t W+\p_\t {\cal K}W= -{1\o \t- \bar
\t}\int_{{\cal M}_6}\overline G\wedge
\O\nonumber\\
 Z_{ij} &=& {\cal D}_i{\cal D}_j W={\kappa_{ijk}\o \int
\O\wedge\overline\O}\int_{{\cal M}_6} G\wedge
\overline\chi^k\label{fzu} \\
Z_{\tau i} &=&{\cal D}_\t{\cal D}_i W=-{1\over \tau -\bar \tau}
\int_{{\cal M}_6} \overline G \wedge \chi_i\cr \quad Z_{\tau \tau}&
=&{\cal D}_\t{\cal D}_\t W= \p_\t F_\t-\G^\t_{\t\t}F_\t+\p_\t{\cal
K}F_\t=0.\nonumber
\end{eqnarray}
As a result the critical condition $\p_I V=0$ can be explicitly
written as
\begin{equation}
\label{critical}
 \left\{
  \begin{array}{ll}
   \int \overline G\wedge \chi_i\int \overline G\wedge
\bar \chi^i +\int \overline G\wedge
\O\int \overline G\wedge \overline \O=0\\\\
\int G\wedge\bar \chi^k\int \overline G\wedge \bar
\chi^i({\kappa_{ijk}\o \int \O\wedge\overline \O}) +\int G\wedge
\chi_j\int \overline G\wedge \overline\O+\int \overline G\wedge
\chi_j\int G\wedge \overline \O=0
  \end{array}
\right.
\end{equation}
After using the Hodge decomposition for $G$
\begin{equation}
\label{decompose}
 G=A\O+A^i\chi_i+\bar B^{\bar i}\bar \chi_{\bar
i}+\bar B\overline \O
\end{equation}
the condition (\ref{critical}) can be further written in the form
\begin{equation}
\left\{
  \begin{array}{ll}
  \int G\wedge \star G=0\\\\
    (B\bar B_k+A\bar A_k)\int \O\wedge\overline
  \O+\kappa_{ijk}A^i B^j=0
\end{array}
\right.
\end{equation}
which are Eq.(\ref{aiii}) and (\ref{condition}). To derive these
equations, we have used the property that the harmonic (2,1) and
(0,3) forms are imaginary self-dual, and the harmonic (1,2) and
(3,0) forms are imaginary anti-self-dual on Calabi-Yau three-fold.

Now we are going to compute the second derivative of scalar
potential by noting that
\begin{equation}
\p_I\p_JV={\cal D}_I{\cal D}_J V,\;\;\;\;\;\p_I\p_{\bar J}V={\cal
D}_I{\cal D}_{\bar J} V
\end{equation}
at the critical point $\p_I V=0$. After a little algebra, the second
derivatives of the scalar potential (\ref{Supergravity2}) are
\begin{eqnarray}
\label{seconderivative2}
\p_I\p_J V&=&e^{\cal K}\left(U_{IJK}\bar F^K +2Z_{IJ}\overline W \right)\nonumber\\
\p_I\p_{\bar J}V&=&e^{\cal K}(U_{\bar JIK}\bar F^K+F_I\bar F_{\bar
J} +Z_{IL}\bar Z_{\bar J\bar K}g^{L\bar K} +g_{I\bar J}|W|^2),
\end{eqnarray}
where $U_{\bar JIK}={\cal D}_{\bar J}{\cal D}_I{\cal D}_K W$. The
above formula can be easily transformed to (\ref{seconderivative})
by using the identity:
\begin{equation}
\label{r} [ {\cal D}_{I},{\cal D}_{\bar J}]F_K=-g_{I\bar J}
F_K+R_{I\bar J K}^{\;\;\;\;\;\;\;L}F_L
\end{equation}

To get expression (\ref{Hessian}), we need to generalize the
definition of $U_{IJK}$ and $Z_{IJ}$ to
\begin{equation}
U_{\a\b\g}={\cal D}_\a{\cal D}_\b{\cal D}_\g W\qquad {\rm and}\qquad
\overline U_{\bar\a\bar \b\bar\g}=\overline{U_{\a\b\g}}
\end{equation}
and
\begin{equation}
Z_{\a\b}={\cal D}_\a{\cal D}_\b W\qquad {\rm and}\qquad \overline
Z_{\bar\a\bar \b}=\overline{Z_{\a\b}},
\end{equation}
where $\alpha$,$\b$,and $\g$ label all coordinates, {\it i.e.} the
axio-dilaton, complex structure moduli and their complex conjugates.
Using the results quoted in the table (\ref{table}), we have
\begin{eqnarray}
U_{ijk}
&=&{\cal D}_i{\cal D}_j{\cal D}_k W={\int G\wedge
\overline \O\o\int \O\wedge \overline\O}\kappa_{ijk}\nonumber\\
U_{ij\t}&=& {\cal D}_i{\cal D}_j{\cal D}_\t W =-{\int \overline
G\wedge \bar
\chi^k\o \int \O\wedge\overline \O}{\kappa_{ijk}\o \t-\bar \t}\label{u1}\\
U_{\bar kij}&=&-{1\o (\int \O\wedge
\overline\O)^2}\kappa_{ij}^{\;\;\;\bar m}\bar \kappa_{\bar k\bar
m\bar n}F^{\bar n}\nonumber
\end{eqnarray}
One consequence of Eq. (\ref{u1}) and Eq. (\ref{fzu}) is
\begin{equation}
\label{u2} \begin{split} & \bar F^{ \t}U_{ij\t}=\bar F^k U_{ijk},\\
& Z_{\bar JI}=g_{I\bar J}W,\\ & Z_{ J\bar I}=0, \\ &  U_{K \bar J
I}=g_{I\bar J}F_K,\\ & U_{\t\t i}=U_{\t\t\t}= U_{\bar K \bar J I}=
U_{\a \bar j\t }=0.\end{split}
\end{equation}
The above expressions are useful to show the equivalence of
(\ref{Hessian}) and (\ref{seconderivative2}).

\section {Appendix B} In this appendix we explicitly show the
appearance of the two superpotentials
\begin{equation}
W=\int G\wedge \O,\qquad {\rm and} \qquad {\widetilde
W}=\int\overline G\wedge \O,
\end{equation}
by dimensional reduction of ten-dimensional supergravity theories.
Our convention is $\e_{01....9}=1$, and
\begin{equation}
\star dx^{m_0}\wedge ...\wedge dx^{m_n}={1\o
(9-n)!}\ep_{m_{n+1}...m_9}^{\;\;\;\;\;\;\;\;\;
\;\;\;\;\;\;\;\;m_0...m_n}dx^{m_{n+1}}\wedge ...\wedge dx^{m_9}
\end{equation}
We take the type IIB effective action (\ref{action}) together with
the local terms are
\begin{equation}
S_{loc}=-\int_{R^4\times \Sigma}d^{p+1}\xi T_p\sqrt{-\hat
G}+\m_p\int_{R^4\times \Sigma} C_{p+1} \label{local}
\end{equation}
To perform the dimensional reduction, we assume that the metric is
independent of external coordinates
\begin{equation}
ds^2=e^{2A(y)}\eta_{\m\n}dx^\m dx^\n+e^{-2A(y)}\tilde g_{mn}(y)dy^m
dy^n\label{metric}
\end{equation}
The Einstein equation is
\begin{equation}
\label{Einstein} R_{MN}=k_{10}^2\bigg(T_{MN}-{1\o 8}g_{MN}T\bigg)
\qquad {\rm with } \qquad T_{MN}=-{2\o\sqrt {-g}}{\d S\o \d g^{MN}},
\end{equation}
The non-compact components of the Einstein equation can be written
as
\begin{equation}
\label{external} R_{\m\n}=\bigg[-{1\o 8\mbox{Im}\t}|G|^2-{1\o
4}e^{-8A}(\p_m\a)^2\bigg]g_{\m\n}+k_{10}^2\bigg(T^{loc}_{\m\n}-{1\o
8}T^{loc}g_{\m\n}\bigg)
\end{equation}
On the other hand, using the metric (\ref{metric}), we obtain
\begin{equation}
R_{\m\n}=-e^{2A}\tilde \nabla^2A g_{\m\n}
\end{equation}
which yields
\begin{equation}
\tilde \nabla^2A ={1\o 8\mbox{Im}\t}e^{-2A}|G|^2+{1\o
4}e^{-10A}|\p_m\a|^2+{1\o 8}k_{10}^2e^{-2A}[T_{m}^{m}-T_\m^\m
]^{loc}.
\end{equation}
This can also be written in the form
\begin{equation}
\tilde \nabla^2e^{4A} ={1\o
2\mbox{Im}\t}e^{2A}|G|^2+e^{-6A}\bigg[(\p_m\a)^2+(\p_m
e^{4A})^2\bigg]+{1\o 2}k_{10}^2e^{2A}\bigg(T_{m}^{m}-T_\m^\m
\bigg)^{loc} \label {nogo1}
\end{equation}

To compute the equation of motion for $C_4$ we only need to consider
a few terms in the action namely
\begin{equation}
{1\o 8{\kappa}_{10}^2}\int \tilde F_{(5)}\wedge \star \tilde
F_{(5)}-{1\o 8i \kappa_{10}^2}\int{C_{(4)}\wedge G\wedge \overline
G\o \mbox{Im}\t}+{\m_p\o 2}\int_{R^4\times \Sigma} C_{p+1}
\end{equation}
The appearance of extra factor ${1\o2}$ is a consequence of the
self-duality of the five form. The Bianchi identity is
\begin{equation}
d\star\tilde F_{(5)}=-{G\wedge \overline G\o
2i\mbox{Im}\t}+2k_{10}^2T_3\r_3^{loc}
\end{equation}
As $\tilde F_{(5)}$ is self-dual, we have
\begin{equation}
\tilde F_{5}=(1+\star)d\a\wedge dx^0\wedge dx^1\wedge dx^2\wedge
dx^3
\end{equation}
and the Bianchi identity becomes
\begin{equation}
\tilde \nabla^2 \a={i\o 12\mbox{Im}\t}e^{2A}G_{mnp}\star\overline
G^{mnp}+2e^{-6A}\p_me^{4A}\p^m\a+2k_{10}^2 T_3\r^{loc}_3
\label{nogo2}
\end{equation}
By summing or subtracting equations (\ref{nogo1})and (\ref{nogo2}),
we get
\begin{equation}
\begin{split}\label{nogo}
\tilde \nabla^2(e^{4A}\pm\a) =& {1\o 2\mbox{Im}\t}e^{2A}|G\mp i\star
G|^2+ e^{-6A}|\p_m\a\pm\p_m e^{4A}|^2\\ & +2
k_{10}^2e^{2A}\bigg({1\o 4}(T_{m}^{m}-T_\m^\m )^{loc}\pm
T_3\r^{loc}_3\bigg).
\end{split}
\end{equation}
The left hand side of the above equation vanishes when integrated
over a compact manifold ${\cal M}_6$. As a result there are two
solutions
\begin{equation}
\begin{split}
& \star_6 G=-iG,\qquad \a=-e^{4A},\qquad  {\rm with}\qquad
\overline
O3,\overline D3\\
& \star_6 G=+iG,\qquad \a=+e^{4A},\qquad {\rm  with}\qquad  O3, D3.
\end{split}
\end{equation}
Notice that we can not have  $O3$ and $\overline D3$ at the same
time.

Using the results above we can perform the dimensional reduction
\begin{equation}
\int d^{10}x\sqrt{-g}{\cal R}=\int d^{4}x\sqrt{-g_4} \int
d^{6}y\sqrt{g_6}\left[-8(\nabla A)^2e^{4A}\right].
\end{equation}
Taking into account the fact the self-duality of the five-form we
get
\begin{equation}
\int d^{10}x\sqrt{-g}{\tilde F^2_{(5)}\o 4}=\int d^{4}x\sqrt{-g_4}
\int d^{6}y\sqrt{g_6}{e^{-4A}\o 2}(\p_m\a)^2
\end{equation}
Since $\a= \mp e^{4A}$, this term gives the same contribution as the
Einstein term
\begin{equation}\label{zzi}
\begin{split}
& \int d^{10}x\sqrt{-g}\left({\cal R}-{\tilde F^2_{(5)}\o 4}\right)
=\int d^{4}x\sqrt{-g_4} \int d^{6}y\sqrt{g_6}\left(-(\p_m
\a)^2e^{4A}\right)\\
& =\int d^{4}x\sqrt{-g_4} \int d^{6}y\sqrt{g_6} \bigg(\mp
(\p_m\a)^2\pm 4\p_m\a\p^m A\bigg)\\ & =\int d^{4}x\sqrt{-g_4} \int
d^{6}y\sqrt{g_6}\bigg(\pm {1\o 12i \mbox{Im}\t}e^{4A}G_{mnp}\star
\overline G^{mnp}\mp 2e^{4A}\kappa_{10}^2T_3\r_3^{loc}\bigg)\\
\end{split}
\end{equation}
Where we have used the Bianchi identity (\ref{nogo2}). The second
term in the last equation of (\ref{zzi}) will cancel the first term
of $S_{loc}$, and the CS term cancels the second term of $S_{loc}$.
At the end, the scalar potential is
\begin{equation}
S_v={1\o 2\kappa_{10}^2}\int d^{4}x\sqrt{-g_4} \int {e^{4A}\o
2\mbox{Im}\t}G\wedge\star_6 (\overline G\pm i\star\overline G)
\end{equation}
From this expression, we can write the scalar potential in the
standard form with
\begin{equation}
{\widetilde W}=\int \overline G\wedge \O,\qquad {\rm or}\qquad
W=\int G\wedge \O
\end{equation}


\begin{thebibliography}{10}


\bibitem{bbw}
  K.~Becker, M.~Becker and J.~Walcher,
  ``Runaway in the Landscape,''
  arXiv:0706.0514 [hep-th].
  %%CITATION = ARXIV:0706.0514;%%


\bibitem{kklt}
S. Kachru, R. Kallosh, A. Linde and S. Trivedi, ``de Sitter vacua in
string theory,'' Phys. Rev. {\bf D68} (2003) 046005
[arXiv:hep-th/0301240].


\bibitem{superpotential}
S.~Gukov,C.~Vafa and E.~Witten,``CFTs from Calabi-Yau Fourfolds,"
Nucl. Phys. {\bf B584}, 69 (2000)[arXiv:hep-th/9906070];
T.~R.~Taylor and C.Vafa,"RR flux on Calabi-Yau and partial
supersymmetry breaking," Phys.Lett. {\bf B474}, 130
(2000)[arXiv:hep-th/9912152]; P.~Mayr,``On supersymmetry breaking in
string theory and its realization in brane," Nucl. Phys. {\bf B593},
99 (2001)[arXiv:hep-th/0003198].


\bibitem{ss}
A. Saltman and E. Silverstein, ``The scale of the No-Scale potential
and de sitter Model Building,'' [arXiv:hep-th/0411271].

\bibitem{gkp}
S.~B.~Giddings, S.~Kachru and J.~Polchinski, ``Hierarchies from
fluxes in string compactifications,'' Phys. Rev. {\bf D66}, 106006
(2002) [arXiv:hep-th/0105097].



\bibitem{denef+douglas}
F.~Denef, M.~R.~Douglas,``Distributions of nonsupersymmetric flux
vacua," [arXiv:hep-th/0411183].



\bibitem{kst}
S.~Kachru, M.~Schulz, S.~P.~Trivedi, ``Moduli stabilization from
fluxes in a simple IIB Orientifold," [arXiv:hep-th/0201028].




\bibitem{fp}
A.~R.~Frey and J.~Polchinski,``$N=3$ warped compactification," Phys.
Rev. {\bf D65} (2002) 126009 [arXiv:hep-th/0201029].


%\cite{Shmakova:1996nz}
\bibitem{Shmakova:1996nz}
  M.~Shmakova,
  ``Calabi-Yau black holes,''
  Phys.\ Rev.\  D {\bf 56}, 540 (1997)
  [arXiv:hep-th/9612076].
  %%CITATION = PHRVA,D56,540;%%


%\cite{Moore:1998pn}
\bibitem{Moore:1998pn}
  G.~W.~Moore,
  ``Arithmetic and attractors,''
  [arXiv:hep-th/9807087].
  %%CITATION = HEP-TH/9807087;%%


\bibitem{attrac} F.~Denef,
  ``Supergravity flows and D-brane stability,''
  JHEP {\bf 0008}, 050 (2000)
  [arXiv:hep-th/0005049].
  %%CITATION = JHEPA,0008,050;%%
  K.~Goldstein, N.~Iizuka, R.~P.~Jena and S.~P.~Trivedi,
  ``Non-supersymmetric attractors,''
  Phys.\ Rev.\  D {\bf 72}, 124021 (2005)
  [arXiv:hep-th/0507096].
  %%CITATION = PHRVA,D72,124021;%%
%\cite{Saraikin:2007jc}
  K.~Saraikin and C.~Vafa,
  ``Non-supersymmetric Black Holes and Topological Strings,''
  [arXiv:hep-th/0703214].
  %%CITATION = HEP-TH/0703214;%%

%\cite{Kallosh:2005ax}
\bibitem{Kallosh:2005ax}
  S.~Ferrara, G.~W.~Gibbons and R.~Kallosh,
  ``Black holes and critical points in moduli space,''
  Nucl.\ Phys.\  B {\bf 500}, 75 (1997)
  [arXiv:hep-th/9702103].
  %%CITATION = NUPHA,B500,75;%%
  R.~Kallosh,
  ``New attractors,''
  JHEP {\bf 0512}, 022 (2005)
  [arXiv:hep-th/0510024].
  %%CITATION = JHEPA,0512,022;%%

\bibitem{Moore:2004fg}
  G.~W.~Moore,
  ``Les Houches lectures on strings and arithmetic,''
  [arXiv:hep-th/0401049].
  %%CITATION = HEP-TH/0401049;%%

\bibitem{candelas}
P.~ Candelas and X. de la Ossa,``Moduli space of Calabi-Yau
Manifolds," Nucl. Phys. {\bf B355}, 455 (1991).

\bibitem{aft}
  L.~Andrianopoli, S.~Ferrara and M.~Trigiante,
  ``Fluxes, supersymmetry breaking and gauged supergravity,''
  [arXiv:hep-th/0307139].
  %%CITATION = HEP-TH/0307139;%%

  %\cite{Lebedev:2006qc}
\bibitem{Lebedev:2006qc}
  O.~Lebedev, V.~Lowen, Y.~Mambrini, H.~P.~Nilles and M.~Ratz,
  ``Metastable vacua in flux compactifications and their phenomenology,''
  JHEP {\bf 0702}, 063 (2007)
  [arXiv:hep-ph/0612035].
  %%CITATION = JHEPA,0702,063;%%

\bibitem{nogo}
 R.~Brustein and S.~P.~de Alwis,
  ``Moduli potentials in string compactifications with fluxes: Mapping the
  %discretuum,''
  Phys.\ Rev.\  D {\bf 69}, 126006 (2004)
  [arXiv:hep-th/0402088].
  %%CITATION = PHRVA,D69,126006;%%
 M.~Gomez-Reino and C.~A.~Scrucca,
  ``Locally stable non-supersymmetric Minkowski vacua in supergravity,''
  JHEP {\bf 0605}, 015 (2006)
  [arXiv:hep-th/0602246].
  %%CITATION = JHEPA,0605,015;%%
 O.~Lebedev, H.~P.~Nilles and M.~Ratz,
  ``de Sitter vacua from matter superpotentials,''
  Phys.\ Lett.\  B {\bf 636}, 126 (2006)
  [arXiv:hep-th/0603047].
  %%CITATION = PHLTA,B636,126;%%



\end{thebibliography}
\end{document}